# The emergence of superconductivity in BaNi$_2$(Ge$_{1-x}$P$_x$)$_2$ at a structural instability

Daigorou Hirai, F. von Rohr*, and R. J. Cava

Department of Chemistry, Princeton University, Princeton, New Jersey 08544, USA

The physical properties and structural evolution of the 122-type solid solution BaNi$_2$(Ge$_{1-x}$P$_x$)$_2$ are reported. The in-plane $X$-$X$ ($X$ = Ge$_{1-x}$P$_x$) dimer formation present in the end member BaNi$_2$Ge$_2$, which results in a structural transition to orthorhombic symmetry, is completely suppressed to zero temperature on P substitution near $x$ = 0.7, and a dome-shape superconducting phase with a maximum $T_c$ = 2.9 K emerges. Clear indications of phonon softening and enhanced electron-phonon coupling are observed at the composition of the structural instability. Our findings show that dimer breaking offers new possibilities as a tuning parameter of superconductivity.

*present address: Physik-Institut, Universität Zürich, Winterthurerstrasse 190, CH-8057 Zürich, Switzerland

Several superconductors with high transition temperatures ($T_c$) are located near structural instabilities, for example, A15 compounds [1], perovskite bismuthates [2], cubic tungsten bronzes [3] and intercalated graphite under high-pressure [4,5]. The instability that leads to the structural phase transition is characterized by phonon softening, and the resulting enhanced electron-phonon coupling is believed to play a significant role for the high $T_c$s. The relationships between structural instabilities and superconductivity have provided much work of interest in the past 50 years.

The tetragonal ThCr$_2$Si$_2$ ("122") structure type is a layered structure with edge-sharing transition metal-metalloid $TX_4$ tetrahedra making $T_2X_2$ layers that alternate with layers of the large $A$ atoms ($A$: alkali metal, alkaline earth metal, rare earth; $T$: transition metal; $X$: metalloid). Previous studies have revealed that some $AT_2X_2$ compounds are located at a structural instability between two different 122 structural types: a collapsed tetragonal (cT) phase and an uncollapsed tetragonal (ucT) phase. [6,7] The structural change from ucT to cT, resulting in significantly reduced ratios of stacking to in-plane lattice parameters (i.e $c/a$) is driven by the formation of a molecule-like $X$-$X$ dimer across the $A$ atom layer. Lattice collapse transitions from the ucT to the cT phase can be controlled by applied physical [8-10] and chemical pressures [11-13], and recent research has demonstrated the usability of chemical bond breaking as a tuning parameter to control physical properties. [14-16]

BaNi$_2$P$_2$ is a ucT 122-type superconductor with a $T_c$ of 2.5 K. [17,18]. BaNi$_2$Ge$_2$, on the other hand, is a 122 type normal metal with a lattice distortion due to Ge-Ge dimer formation, but not of the cT-ucT lattice collapse type. It crystallizes in an ordinary ucT ThCr$_2$Si$_2$-type structure at high temperature, but exhibits a structural transition to an orthorhombic phase (space group *Pnma*) on cooling below 753 K. [19] The important consequence of the structural transition is that equal Ge-Ge distances in the Ni$_2$Ge$_2$ layers at high temperature (3.61 Å at 783 K) are split into one short distance (2.93 Å at 753 K) and two much longer distances (3.69 Å and 3.72 Å at 753K) below the phase transition, as shown in Fig. 1(a). This structural transition is the result of in-plane Ge-Ge dimer formation in the Ni$_2$Ge$_2$ layer. This unusual structural instability is driven by the strong tendency toward Ge-Ge bonding in 122 type structures as a consequence of the electron count [6] even though, due to the presence of the large Ba ion, the layers are too far apart in $BaNi_2Ge_2$ for the usual $X$-$X$ bond across the $A$ layer seen in the cT phases.

In this letter, we demonstrate the systematic suppression of the in-plane dimer phase of $BaNi_2Ge_2$ by P doping. A dome-shape superconducting phase emerges at the structural instability where the dimer phase is suppressed and an ucT phase appears. The enhanced specific heat jump at the superconducting transition at the critical border of instability of the dimer phase and the ucT phase strongly suggests that phonon-softening induced by dimer breaking is responsible for the enhanced $T_c$. Our finding demonstrates that dimer breaking offers a new possibility as a tuning parameter of superconductivity.

Conventional solid-state reactions were employed to synthesize the polycrystalline samples. As the first step, $BaNi_2P_2$ and $BaNi_2Ge_2$ were made by mixing elemental Ba (99%), Ni (99.9%) and P (red, 99%) or Ge (99.9999%) at the ratio of 1.05 : 2 : 2 in an argon filled glove box. The mixed powder was placed in an alumina crucible and sealed in an evacuated quartz tube, and then sintered initially at 550 °C for 10 hr and 800 °C for 48 hr. Then, $BaNi_2P_2$ and $BaNi_2Ge_2$ were mixed as $BaNi_2(Ge_{1-x}P_x)_2$ for $0 < x < 1$, heated at 800 °C for 24 hr. The sintered pellet was reground, repelletized and sintered again at 800 °C for 24 hr.

All products were initially characterized by laboratory powder X-ray diffraction (XRD, Bruker D8, Cu K$\alpha$ radiation). Selected samples were characterized with high resolution synchrotron powder diffraction at beamline 11-BM at the Advanced Photon Source (APS) at Argonne National Laboratory using an average wavelength of 0.41308 Å. The structures were refined by the Rietveld method using the FullProf program. [20]

Magnetic susceptibility, resistivity and heat capacity measurements were performed by a Magnetic Properties Measurement System (MPMS: Quantum Design) and a Physical Property Measurement System (PPMS: Quantum Design). Specific-heat measurements down to 0.4 K were performed in a PPMS equipped with a $^3$He insert.

The structural instability that results from in-plane dimer breaking, which can be recognized as the structural transition from an orthorhombic phase to a tetragonal phase, was clearly observed between $x = 0.2$ and 0.4 in the room temperature XRD patterns. As shown in the inset of Fig. 1, the 112 and 011 peaks in the orthorhombic cell at low P content become a single 112 peak in the tetragonal cell above a P content of $x = 0.3$. This disappearance of peak splitting indicates that the distortion of the $Ni_2X_2$ ($X = Ge_{1-x}P_x$) layers originating from the $X$-$X$ dimer formation is relaxed and that the $Ni_2X_2$ lattice becomes a regular square.

The XRD pattern for x = 0.3 can be refined with a mixture of tetragonal $BaNi_2P_2$-like and orthorhombic $BaNi_2Ge_2$ -like phases with a ratio of 6∶4. There is no indication of chemical phase separation into $BaNi_2P_2$-like and $BaNi_2Ge_2$ -like phases, except for $x = 0.3$, in $BaNi_2(Ge_{1-x}P_x)_2$, and the XRD patterns of $BaNi_2(Ge_{1-x}P_x)_2$ with $x$ below and above 0.3 were well refined with orthorhombic (*Pnma*) and tetragonal cells (*I4/mmm*), respectively. The systematic change in the room temperature XRD pattern indicates that the dimer formation at 753 K in $BaNi_2Ge_2$ [19] is driven down in temperature by P substitution and straddles room temperature at a P content of $x = 0.3$.

Figure 1 shows the room temperature lattice parameters as a function of P concentration for $BaNi_2(Ge_{1-x}P_x)_2$ determined by the refinement of the XRD data. The *b*-axis (*c*-axis in the tetragonal phase), a measure of unit-cell perpendicular to the $Ni_2X_2$ layers, decreases monotonically, but nonlinearly, with P content $x$. On the other hand, the *a* and *c*-axes, (*a*-axis in the tetragonal phase), measures of the $Ni_2X_2$ in-plane dimensions, increase. This anisotropic change in lattice parameters with increasing P content is in striking contrast to what is seen in the highly analogous conventional solid solution $LaNi_2(Ge_{1-x}P_x)_2$ where they decrease monotonically with increasing P content [21] due to the simple chemical pressure effect that occurs when substituting P (atomic radius: 1.00 Å) for Ge (atomic radius: 1.25 Å). [22] When the Ge is totally substituted by P, the crystal structure of $BaNi_2P_2$ is in a ucT phase with a large ratio of stacking to in-plane lattice parameters (*c/a*) of 3.0, consistent with previous reports. [23]

As the transition temperature of in-plane dimer breaking ($T_s$) is suppressed with increasing P content, $T_s$ comes into the temperature range of our resistivity measurements and is clearly observed above $x = 0.3$, as shown in Fig. 2. All the $BaNi_2(Ge_{1-x}P_x)_2$ samples have resistivities below 1 mΩcm at 300 K, indicating the metallic nature of these compounds. The resistivities of both of the end members $BaNi_2Ge_2$ ($x = 0$) and $BaNi_2P_2$ ($x = 1$) show a monotonic decrease on cooling without any anomaly. In the intermediate doping range of $x = 0.3 \sim 0.6$, however, the resistivity data show a local maximum accompanied by hysteresis. The clear hysteresis in resistivity data indicates that the dimer breaking transition is of first order. From $x = 0.3$ to $x = 0.6$ the peak position systematically decreases in temperature and the hysteresis smears out. Finally, the anomaly completely disappears above P contents of $x = 0.7$ and normal metallic behavior is recovered.

When the structural transition observed through the anomaly in the resistivity data of BaNi$_2$(Ge$_{1-x}$P$_x$)$_2$ is completely suppressed at a P content $x = 0.7$, a superconducting transition is observed in low-temperature magnetization data. Figure 3(a) shows that a large diamagnetic signal hallmarking bulk superconductivity is observed in the samples with a P content of $x = 0.7 \sim 1.0$. The $T_c$ reaches a maximum of about 2.9 K for $x = 0.8$ and then decreases slightly for higher P content. The $T_c = 2.5$ K for the $x = 1$ compound BaNi$_2$P$_2$ is consistent with the previous report for a single crystal. [18] Further investigation of the superconducting phase down to 0.4 K was performed by heat capacity measurement. A large heat capacity jump corresponding to the superconducting transition is clearly observed in BaNi$_2$(Ge$_{1-x}$P$_x$)$_2$ with $x \geq 0.7$ (Fig. 3(b)). Superconducting transitions were also observed for the lower P contents of $x = 0.4 \sim 0.6$. Below $x = 0.3$, superconductivity is no longer observed above our lowest measurement temperature of 0.4 K. All the samples show very sharp transitions indicating that they are highly homogeneous.

Detailed analysis of the low temperature specific heat data shows clear indication of the criticality of the dimer-breaking structural instability. The normal-state heat capacity data under an applied field that suppresses the superconductivity can be well fitted by the equation $C/T = \gamma + \beta T^2$, where $\gamma$ is the electronic specific heat coefficient and $\beta$ is the phonon contribution. As the P content increases, $\gamma$ increases monotonically from BaNi$_2$Ge$_2$ to BaNi$_2$P$_2$, while the Debye temperature ($\Theta_D$) calculated from $\beta$ has a minimum around the structural phase instability, as shown in Fig 4(a,b). In contrast, both $\gamma$ and $\beta$ of LaNi$_2$(Ge$_{1-x}$P$_x$)$_2$, where the end members LaNi$_2$Ge$_2$ and LaNi$_2$P$_2$ crystalize in the same cT structure type, decrease monotonically with increasing P content. [21]

The criticality can be more clearly observed in the normalized specific heat jump at the superconducting transition ($\Delta C/\gamma T_c$), as shown in Fig. 4(b). The value of $\Delta C/\gamma T_c$ exhibits a clear maximum centered at the composition boundary between the dimer and ucT phases. The Fermi surface and transport properties of the BaNi$_2$P$_2$ superconductor are distinct from those of the iron-pnictide superconductors [18,24] and BaNi$_2$P$_2$ is considered to be a phonon-mediated conventional superconductor. Considering that $\Delta C/\gamma T_c$ can be underestimated due to non-superconducting contributions to $\gamma$ and the broadness of superconducting transition, the value of $\Delta C/\gamma T_c \sim 1.1$ for BaNi$_2$P$_2$ agrees reasonably with BCS weak coupling value of 1.43. At $x = 0.8$, however, $\Delta C/\gamma T_c$ increases by 50%, reaching 1.63 and suggesting strong electron-phonon coupling.

As summarized in the electronic phase diagram (Fig.4(c)), the emergence of the superconducting phase in BaNi$_2$(Ge$_{1-x}$P$_x$)$_2$ is strongly correlated with the structural instability caused by the in-plane dimer breaking transition. When the P content $x$ increases, the room temperature crystal structure changes from orthorhombic to tetragonal at $x = 0.3$, and the anomaly in resistivity data indicative of the structural phase transition is driven down in temperature. A superconducting phase emerges at $x = 0.5$, and $T_c$ abruptly increases with the dimer phase suppressed; finally $T_c$ reaches a maximum at $x = 0.8$ when the dimer is fully broken.

Electronic-structure calculations show that the valence band around the Fermi level in BaNi$_2$P$_2$ is formed predominantly by a strongly hybridized transition-metal $d$ and metalloid $p$ states. [25] With the presence of covalent Ge-Ge in-plane bonding, the Fermi level for BaNi$_2$Ge$_2$ must be located between the bonding orbitals ($\sigma$) and anti-bonding orbitals ($\sigma^*$) of the Ge-Ge dimer. The dimer is absent in BaNi$_2$P$_2$ due to the higher electron count, which results in occupancy of the antibonding $\sigma^*$ dimer band. Therefore, the structural instability in BaNi$_2$(Ge$_{1-x}$P$_x$)$_2$ is brought about by doping electrons into the $X$-$X$ dimer anti-bonding orbital. [6] This complements our recent research on SrCo$_2$(Ge$_{1-x}$P$_x$)$_2$ [11] and LaCo$_2$(Ge$_{1-x}$P$_x$)$_2$ [13], which revealed that a conventional lattice collapse transition, a result of interlayer $X$-$X$ dimer formation, can be controlled by P doping of germanides.

In contrast to the gradual increase in $\gamma$, clear criticality is observed in both the Debye temperatures and specific heat jumps at $T_c$ in the BaNi$_2$(Ge$_{1-x}$P$_x$)$_2$ system. These observations lead us to propose that enhanced electron-phonon coupling due to softening phonons is responsible for the enhancement of $T_c$ at the boundary of the dimer breaking structural instability. An enhancement of $T_c$ at the boundary of two different structures has been reported, for example, in tellurium under high pressure, where $T_c$ jumps from 2.5 K to 7.4 K. [26] First-principles calculations have pointed out that the jump in $T_c$ in that case can be explained by the enhancement of electron-phonon interaction. [27] Recently, Kudo *et al.*, found a discontinuous increase in $T_c$ from 0.6 K to 3.3 K in BaNi$_2$(As$_{1-x}$P$_x$)$_2$ at the boundary of a triclinic lower P content phase and a tetragonal high P content phase. [28] A similar reduction of Debye temperature and enhanced $\Delta C/\gamma T_c$ is observed in that system. It is interesting to note that $T_c$ shows a step-wise increase in both Te and BaNi$_2$(As$_{1-x}$P$_x$)$_2$ at the phase boundary, in contrast to the dome-shape superconducting phase in BaNi$_2$(Ge$_{1-x}$P$_x$)$_2$. Further theoretical and experimental investigation, especially on soft phonon modes in 122 structure types, will be useful to clarify the difference between these systems and to look at the possibilities for achieving superconductivity with higher $T_c$'s. The strong correlation between the structural transition and superconductivity BaNi$_2$(Ge$_{1-x}$P$_x$)$_2$ gives rise to a novel opportunity for employing dimer breaking as a control parameter of superconductivity.

The authors thank B. Toby, L. Ribaud and M. Suchomel at beam line 11-BM at the APS for their excellent diffraction data, and L. M. Schoop for helpful discussions. This work was supported by the US DOE office of Basic Energy Sciences, grant DE-FG02-45706. Use of the APS at Argonne National Laboratory was supported by the U. S. DOE, Office of Science, Office of Basic Energy Sciences, under Contract No. DE-AC02-06CH11357.

## Figure Captions

Fig. 1 (Color online) (a) A comparison of the P and Ge atom planes in the crystal structures of BaNi$_2$P$_2$ (*I4/mmm*) and BaNi$_2$Ge$_2$ (*Pnma*). Dotted squares represent unit cells. The ovals indicate the shortest Ge-Ge bonds in the Ni$_2$Ge$_2$ layer. (b) The P content dependence of the in-plane and perpendicular to plane lattice parameters of BaNi$_2$(Ge$_{1-x}$P$_x$)$_2$. The inset is an enlarged XRD pattern showing the change in the crystal symmetry with P content. Bold lines and dots represent synchrotron and laboratory XRD patterns, respectively.

Fig. 2 (Color online) The temperature dependence of the electrical resistivity of BaNi$_2$(Ge$_{1-x}$P$_x$)$_2$ with various P concentrations. Each data set is normalized by the resistivity value at 350 K and shifted to avoid overlap. Arrows indicate the low temperature end points of the hysteresis.

Fig. 3 (Color online) The temperature dependence of the dc-magnetization (*M/H*) under applied magnetic field of 10 Oe and the electronic part of the specific heat divided by temperature ($C_e/T$) for BaNi$_2$(Ge$_{1-x}$P$_x$)$_2$

Fig. 4 (Color online) The P content dependence of (a) The electronic specific heat coefficients ($\gamma$), (b) The Debye temperatures ($\Theta_D$) obtained from low temperature fits of the specific heat data and the normalized specific heat jumps at $T_c$ ($\Delta C/\gamma T_c$). Error bars for $\gamma$ are less than the point size. (c) The electronic phase diagram (logarithmic *T*) of BaNi$_2$(Ge$_{1-x}$P$_x$)$_2$. Filled squares represent superconducting transition temperatures ($T_c$) determined from heat capacity measurement. Filled and open triangles are the structural transition temperatures ($T_s$) determined by powder XRD data and obtained from ref. [10], respectively. Open squares and circles represent respectively the $T_s$ upon cooling and heating determined from the resistivity data.

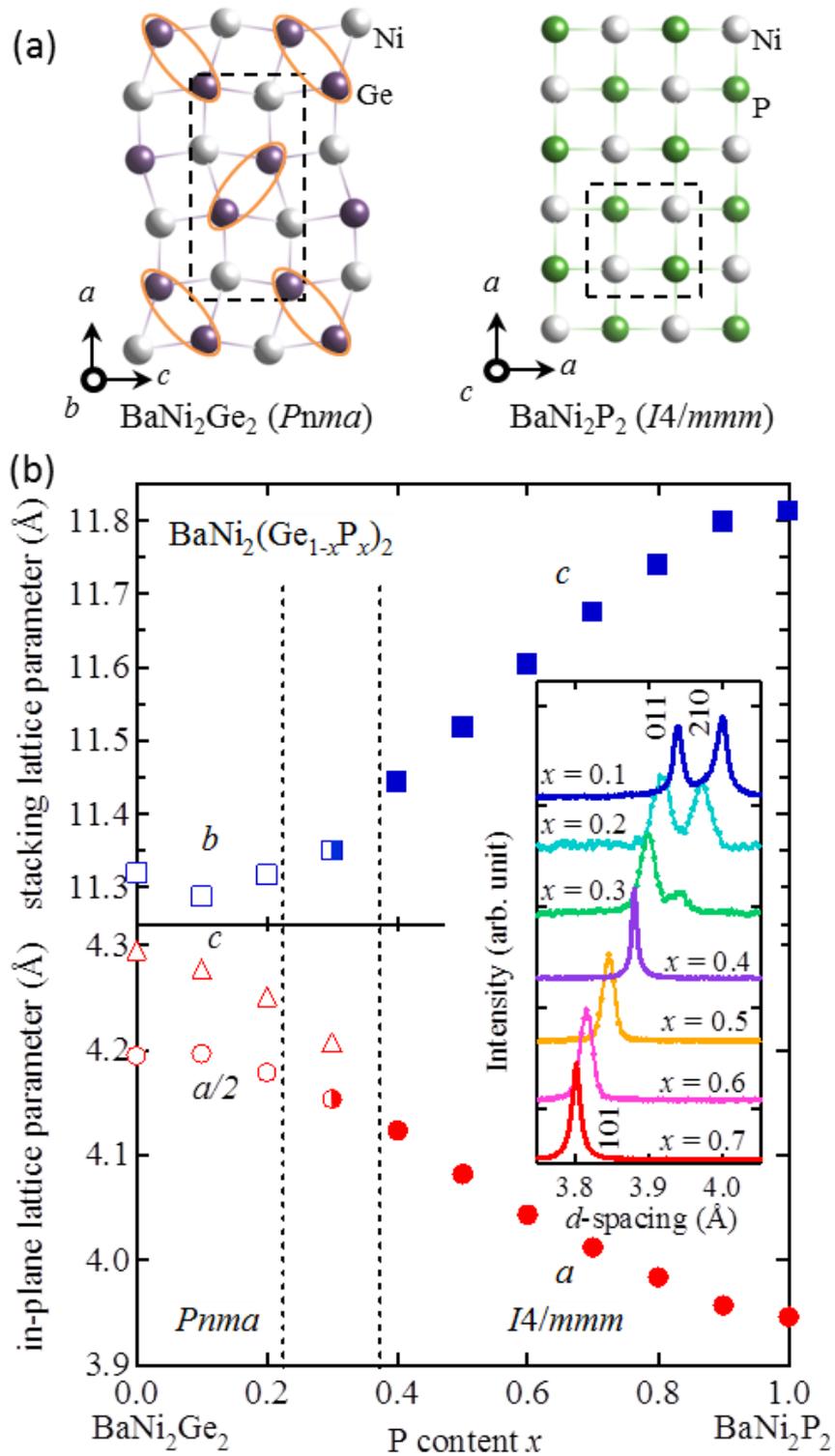

Fig. 1

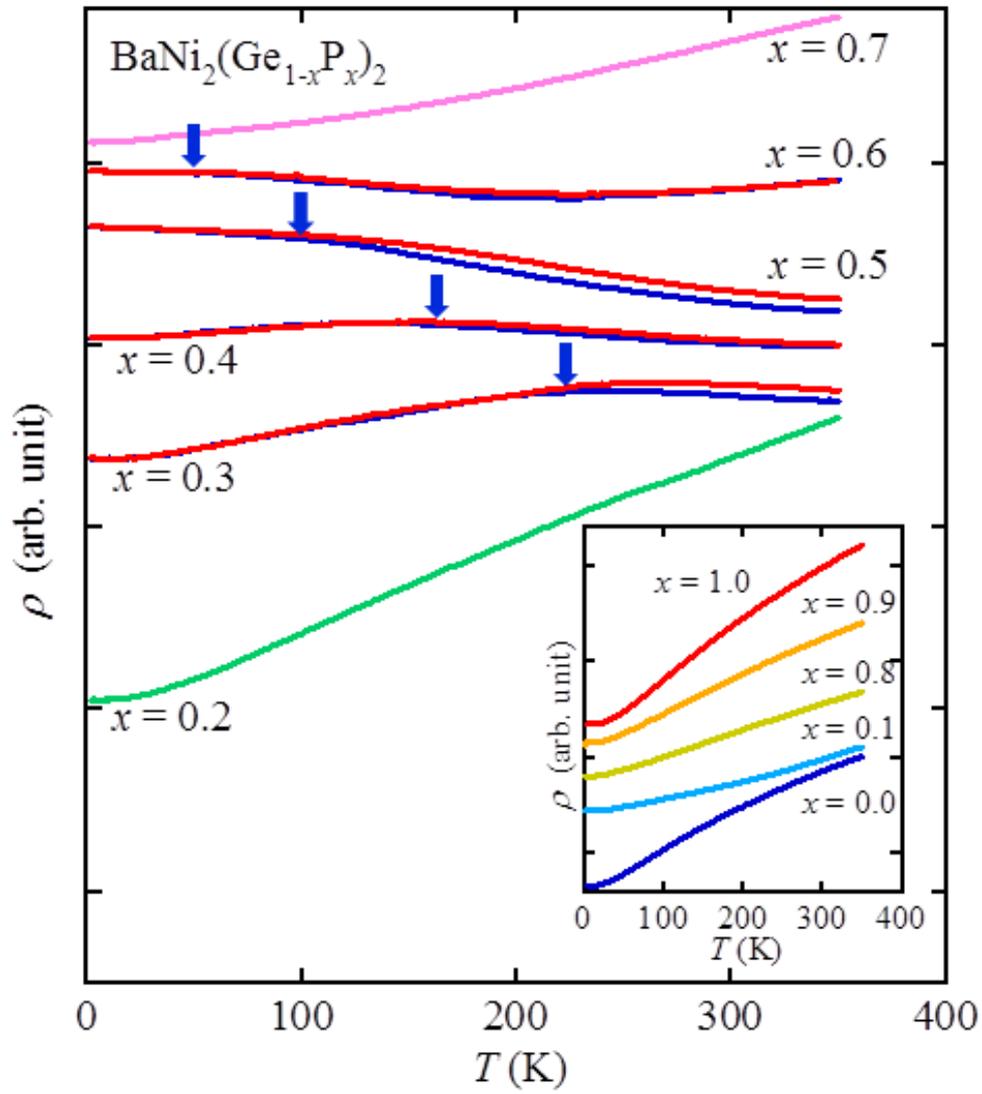

Fig. 2

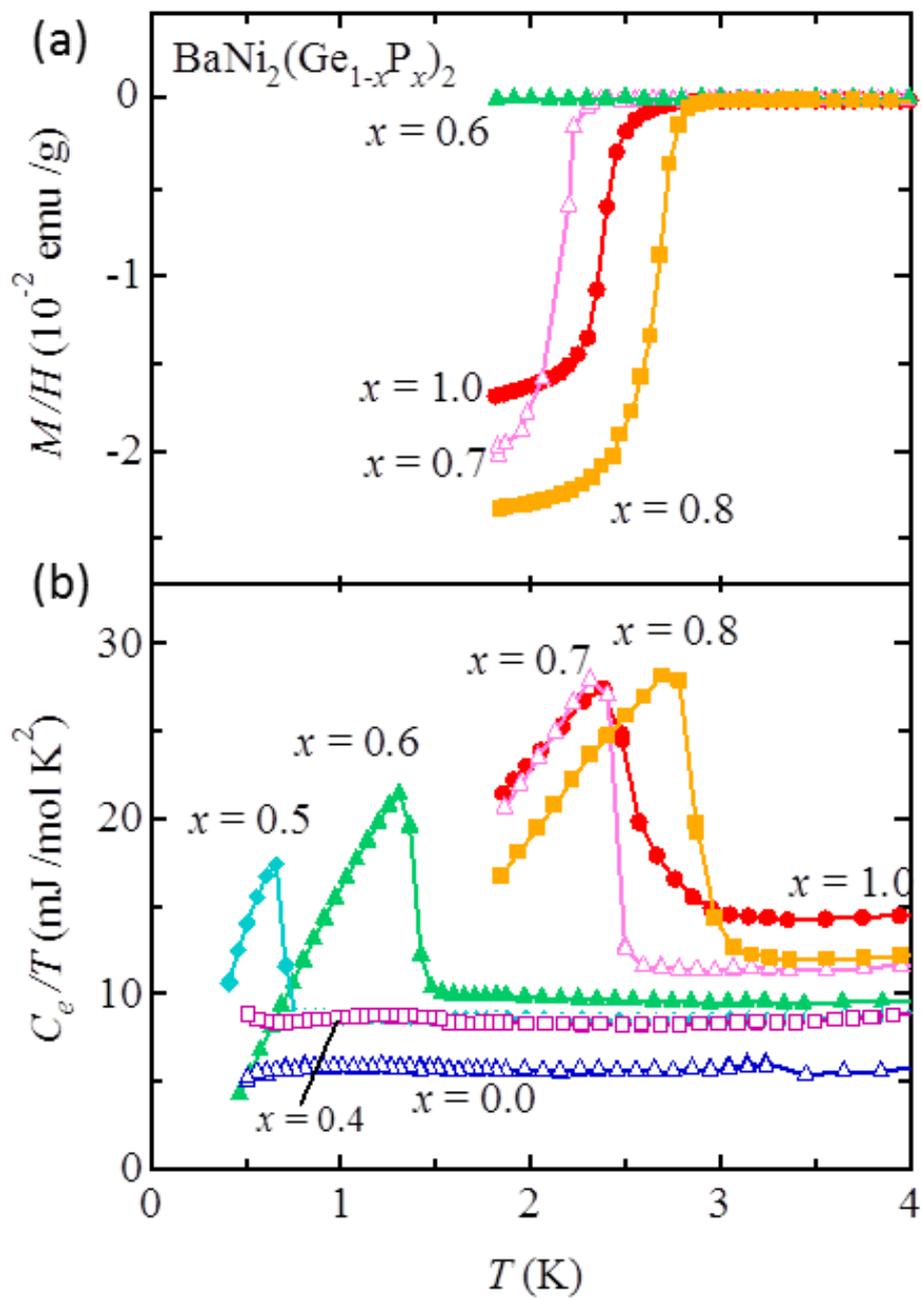

Fig. 3

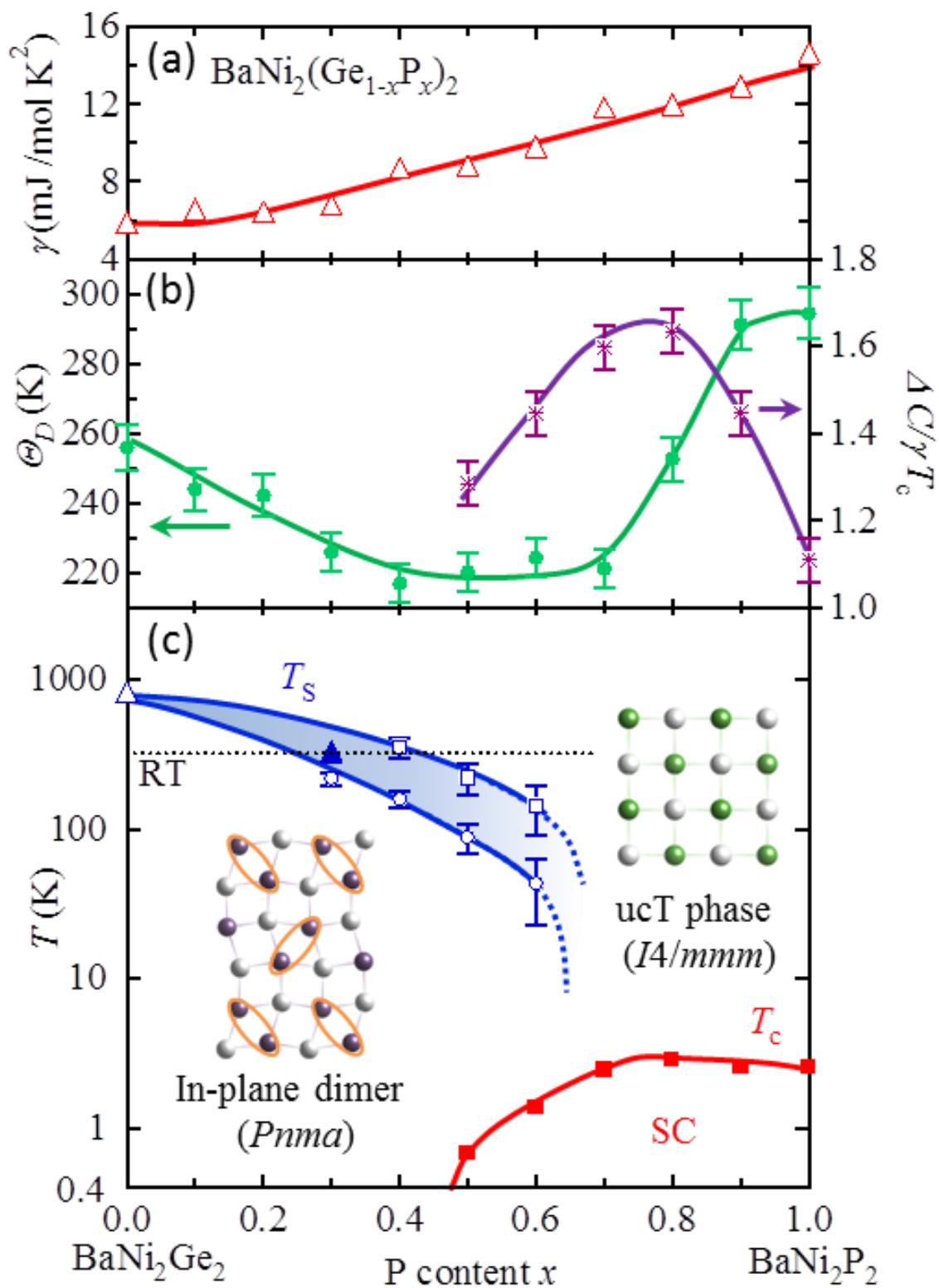

Fig. 4